\documentclass[prd,onecolumn, nofootinbib, 11pt,]{revtex4}  
\usepackage{graphicx}
\usepackage{amsmath,braket}
\usepackage{amsfonts}
\usepackage{amssymb}
\usepackage{bm}
\usepackage{appendix}
\usepackage{mathtools}
\usepackage{comment}
\usepackage{bbold}
\usepackage{color}
\usepackage{slashed}
\usepackage[hyperindex=true, citecolor=green]{hyperref}
\usepackage{subfigure}
\usepackage{setspace}
\usepackage{enumitem}
\usepackage{longtable}
\usepackage{wasysym}
\usepackage[usenames,dvipsnames]{xcolor}
\usepackage{bm}
\usepackage{multirow}
\usepackage{changepage}
\usepackage{kantlipsum}
\usepackage{mathtools}
\usepackage{yfonts}
\usepackage{mathrsfs}
\usepackage[letterpaper, margin=1in]{geometry}
\usepackage{graphicx}

\usepackage{tikz}

\usetikzlibrary{decorations.pathmorphing,arrows.meta,bending}

\pdfoutput=1

\newcommand{\e}{\, .}
\newcommand{\ee}{\, ,}

\begin{document}

\singlespacing

\title{\Large Continuity and Semileptonic $B_{(s)}\rightarrow D_{(s)}$ Form Factors}

\author{Andrew Kobach}
\affiliation{Physics Department, University of California, San Diego, La Jolla, CA 92093, USA}

\date{\today}

\begin{abstract}
Small changes in the masses of massive external scattering states should correspond to small changes in the non-perturbative parameterization of form factors in quantum field theory, as long as the relevant energy range is not near strong deformations.  Here, the definition of ``small'' is investigated and applied to $SU(3)$ breaking in semileptonic $B_{(s)}\rightarrow D_{(s)}$ transitions.    
When unitarity and analyticity are imposed, the differences in the form factors for semileptonic $B\rightarrow D$ versus $B_s\rightarrow D_s$ decays are found to be within $\mathcal{O}(1\%)$ over the entire kinematic range, not just at zero recoil, which is consistent with results from lattice calculations and differs from the expectation using HQET alone.  
\end{abstract}

\maketitle

\section{\normalsize Introduction}
\label{intro}

\singlespacing

Matrix elements of local operators can be decomposed in to a set of non-perturbative scalar functions, called form factors.  These form factors have served as a way to directly compare experimental measurements and non-perturbative theoretical predictions.  
If the masses of the external scattering states are varied by an infinitesimal amount, the form factors should also vary by an infinitesimal amount.  This assumption of continuity is common, though not always explicitly stated.  For example, the  form factors for $\bar{B}^0 \rightarrow D^{+(*)} \ell \overline{\nu}$ are, to a very good approximation, the same as those for $\bar{B}^- \rightarrow D^{0(*)} \ell \overline{\nu}$, since $(M_{B^0} - M_{B^\pm})/\Lambda_\text{QCD} \ll 1$, $(M_{D^0} - M_{D^\pm})/\Lambda_\text{QCD} \ll 1$,  and $\alpha_\text{QED} /4\pi \ll 1$.  Generally put, the interactions of a theory have a nominal energy scale $\Lambda$, and the form factors for scattering processes in the theory will undergo small changes if the masses of the external states change by a small amount compared to $\Lambda$.

If the masses of any external scattering state change by an amount $\varepsilon$, i.e., $M\mapsto M+\varepsilon$, how large in value can $\varepsilon$ be until the form factors begin to change significantly?  $B_{(s)}$ and $D_{(s)}$ mesons are suitable to this analysis, since $\varepsilon \simeq M_{B_s} - M_B \simeq M_{D_s} - M_D$. 
For example, consider purely leptonic decays  $B_{(s)}\rightarrow \ell \nu$ and $D_{(s)}\rightarrow \ell \nu$, with associated decay constants $f_{B_{(s)}}$ and $f_{D_{(s)}}$, respectively.\footnote{Non-perturbative decay constants can be thought of as a form factor sampled at a singular point in momentum.} Recent lattice  measurements are $f_{B_s}/f_B \simeq 1.22$ and $f_{D_s}/f_D \simeq 1.18$~\cite{Bazavov:2017lyh},  which is consistent with the estimate that these ratios scale like $\varepsilon/ \Lambda_\text{QCD}$, as in chiral perturbation theory~\cite{Grinstein:1992qt,Grinstein:1993ys}, and not $\varepsilon/M$, since the masses of the $B$ and $D$ are not similar, i.e., $M_D/M_B \simeq 0.35$. 
Turning to the differences in the form factors for $B \rightarrow D^{(*)} \ell \nu$ versus $B_s \rightarrow D_s^{(*)} \ell \nu$, one might again expect that they do differ by order $\varepsilon/\Lambda_\text{QCD} \sim 10\% - 50\%$.  However, this turns out not to be the case.  In the limit $M\gg \Lambda_\text{QCD}$, heavy particle effective field theory (HQET) predicts that the form factors would differ by the scale $m_s/M$ at zero recoil, and $m_s/\Lambda_\text{QCD}$ away from zero recoil in $q^2$ space.  Recent results from the lattice observe that only the former is true, i.e., that the effect of the valence quark scales like $m_s/M \sim \mathcal{O}(1\%)$ at zero recoil.  In fact, the lattice observes that the form factors associated with the semileptonic decays $B \rightarrow D$ versus the analogous ones for $B_s \rightarrow D_s$ differ from each other at the level of $\mathcal{O}(1\%)$ {\it over the entire kinematic range~\cite{Bailey:2012rr, Bailey:2014tva, Harrison:2017fmw, Monahan:2017uby, McLean:2019qcx}.}   Currently, there is no  explanation as to why this is in the literature.    A quantitative estimate for how form factors change as the masses of the external states are varied is the purpose of this work.  

The discussion will focus on matrix elements between single-particle momentum  eigenstates.  The transition between single-particle states is particularly simple, since there is only a single kinematic factor on which the form factors depend.  More general scenarios are straightforward to consider.  
The expectation of continuity is discussed in Section~\ref{formfactors}, and its effects, combined with the constraints from analyticity and unitarity, are illustrated for semileptonic $B$ decays in Section~\ref{Bdecays}.  It is estimated that the form factors for ${B} \rightarrow D$ versus ${B}_s \rightarrow D_s$ have, to a good approximation, the same shape over the entire kinematic range, not just at zero recoil, and differ only in normalization by a few percent, in agreement with lattice calculations.

\section{\normalsize Form Factors and Continuity}
\label{formfactors}

The matrix element between single particle momentum eigenstates can be decomposed as follows:
\begin{equation}
\label{matrixelement}
\bra{X_f(p',s')} \mathcal{O}^{\{ \mu_n \} }(q) \ket{X_i(p,s)} = \sum_k ~ f_k(q^2)~T_k^{\{ \mu_n \} }(p, p', s, s') \ee 
\end{equation}
where $p - p' = q$, the sum over $k$ is finite, $s$ and $s'$ signify spin degrees of freedom, and $\mathcal{O}^{\{ \mu_n \}}$ is some operator with a set of Lorentz indices $\{ \mu_n \}$ inserting momentum $q$ into the system.  The functions $f_k(q^2)$ are a set of $k$ dimensionless, scalar functions called form factors.  The known functions $T_k^{\{ \mu_n \} }(p, p', s, s')$ inherit the Lorentz structure of the matrix element and, in general, depend on the spin degrees of freedom. Here, we will be considering the region of parameter space where $M_i$ and $M_f$ are both always nonzero.  
The states $X_i$ and $X_f$ are taken to be well-approximated as being on-shell, with masses $M_i$ and $M_f$, respectively.

Considering the expansion of the form factor as a function of $q^2$ around the point $q_0^2$: 
\begin{eqnarray}
f_k(q^2) = \sum_{n=0}^\infty b_n (q^2-q^2_0)^n \ee
\end{eqnarray}
the coefficients $b_n$ can carry dimensions of mass.\footnote{Their values will be bounded due to unitarity, and the exact details regarding such a unitarity bound are discussed for a specific example in Section.~\ref{Bdecays}}  As the masses of the external states, $M_i$ and $M_f$, are changed, i.e., $M_i \mapsto M'_i$ and $M_f \mapsto M'_f$, leaving all other quantum numbers of $X_i$ and $X_f$ the same, one will have a new form factor, $f'_k(q^2)$:
\begin{eqnarray}
f'_k(q^2) = \sum_{n=0}^\infty d_n (q^2-q^2_0)^n \e
\end{eqnarray}
A way to relate $b_n$ and $d_n$ is via dimensionless scalar functions $\mathcal{F}_n$ that depend on $M_i^{(')}$, $M_f^{(')}$, and $\Lambda$:
\begin{equation}
\label{dn}
\frac{d_n}{b_n} =  1 + \mathcal{F}_n \e
\end{equation}
As defined, $\mathcal{F}_n$ is equal to zero when $M'_i = M_i$ and $M'_f = M_f$. Here, we are interested in how $\mathcal{F}$ scales with small variances in the masses:~$M_i' = M_i + \varepsilon$ and $M'_f  = M_f + \delta$, while the remaining quantum numbers remain unchanged.

There are a few limiting cases to consider.  First, if the masses of the external states are much lower than the nominal scale of the underlying interactions, $M\ll \Lambda$, then, 
in general, $\mathcal{F}_n$ then can be approximated as the following, to first order in $\varepsilon$ and $\delta$:
\begin{eqnarray}
\label{contform1}
\mathcal{F}_n \simeq  A_n \frac{\varepsilon}{\Lambda} + B_n \frac{\delta}{\Lambda}  + \mathcal{O}\left( \varepsilon^2, \delta^2, \varepsilon \delta \right) \e
\end{eqnarray}
This would be the expected behavior, for example, in chiral perturbation theory, containing only up and down quarks.  The  assumption of continuity is that this Taylor expansion in Eq.~\eqref{contform1} is not only possible, but it is also useful, i.e., $A_n$ and $B_n$, which can depend on $M_i$, $M_f$, and $\Lambda$, are constants  that are not $\gg \mathcal{O}(1)$.  
This is likely to occur when the range of physical $q^2$ is not near any strong deformations in the theory, where the form factor can vary significantly over a small range of $q^2$, e.g., poles, branch cuts, or regions in the theory not near any ``states,'' but still exhibit significant variations.\footnote{Non-analytic features are kinds of strong deformations, but one need not make the assertion that the only source of high variation in  quantum field theories are only due to non-analytic features.}
On the other hand, if $M \sim \Lambda$, there is only one energy scale, and
\begin{eqnarray}
\label{contform2}
\mathcal{F}_n \simeq  A_n \frac{\varepsilon}{\Lambda} + B_n \frac{\delta}{\Lambda} + C_n \frac{\varepsilon}{M} + D_n \frac{\delta}{M}  + \mathcal{O}\left( \varepsilon^2, \delta^2, \varepsilon \delta \right) \e
\end{eqnarray}
Again, the assumption of continuity can be made, where none of the above coefficients are particularly large.  
If $M_i$ and $M_f$ are greater than $\Lambda$, then a different expectation can be made about the scaling of $\mathcal{F}$.  The rest of this section will explore this statement.

We will proceed without utilizing a Lagrangian, and instead relying solely on scaling arguments as a function of the masses of the external states.  This is similar in spirit to the methods used in Ref.~\cite{Shifman:1987rj,Isgur:1989vq, Isgur:1989ed}, before the discovery of the HQET Lagrangian.  The final results in this section are, unsurprisingly, consistent with the full effective theory~\cite{Manohar:2000dt}.  
However, to our knowledge, the particular argument presented here has not yet been illustrated in the literature, so a detailed derivation is discussed in Appendix~\ref{app} in order to remain self-contained. 

Making the small differences in masses explicit, let $M'_i = M_i + \varepsilon$, and $M'_f = M_f + \delta$, and the ratio the form factors in Eq.~\eqref{dn} scales as follows, where
\small
\begin{eqnarray}
\label{ratio6}
\mathcal{F}_n  &\simeq&  a_1 \frac{\varepsilon}{M_i} + a_2 \frac{\delta}{M_f} + a_3\big(1+\mathcal{\chi}_1(v\cdot v')\big) \left( \frac{\Lambda}{M_i} \frac{\varepsilon}{M_i} + a_4 \frac{\Lambda}{M_f} \frac{\delta}{M_f} \right) \nonumber \\
&& ~~ + ~ \mathcal{\chi}_2(v\cdot v')  \left( \frac{\varepsilon}{\Lambda} + a_5 \frac{\delta}{\Lambda} \right) \left(1 + a_6 \frac{\Lambda}{M_i} + a_7 \frac{\Lambda}{M_f} \right) +  \mathcal{O}\left(\frac{\Lambda^2}{M^2},~ \frac{\varepsilon^2}{M^2},~ \frac{\delta^2}{M^2} \right) \e
\end{eqnarray}
\normalsize
The functions $\mathcal{\chi}_1(v\cdot v')$ and $\mathcal{\chi}_2(v\cdot v')$ are zero when $v\cdot v' = 1$, and $a_{1-7}$ are dimensionless constants, all of which are assumed not to be $\gg \mathcal{O}(1)$, according to continuity.  Of course, the number and location of functions such as $\mathcal{\chi}_1(v\cdot v')$ and $\mathcal{\chi}_2(v\cdot v')$ are not unique.  
Note that Eq.~\eqref{ratio6} describes the behavior of the scaling; one cannot further take $M\rightarrow \infty$. 
The introduction of the factors $a_1$ and $a_2$ account for the possibility of radiative corrections to the ratio on the left-hand size of Eq.~\eqref{ratio6}.  Importantly, such radiative corrections can only scale like $\varepsilon/M$ or $\delta/M$, and not $\varepsilon/\Lambda$ or $\delta/\Lambda$, since, generically speaking, such corrections have a perturbative origin, coming from matching the the effective theory to the full theory, and do not depend on $\Lambda$. 

Eq.~\eqref{ratio6} makes the prediction that at zero recoil, the ratio scales like $\varepsilon/M_i$ and $\delta/M_f$, not like $\varepsilon/\Lambda$ or $\delta/\Lambda$.  Away from zero recoil, corrections to the ratio can scale like $\varepsilon/\Lambda$ or $\delta/\Lambda$.  Because the transition matrix elements are expected to scale as in Eq.~\eqref{ratio6} when $M>\Lambda$, then the form factors associated with those matrix elements, as defined in Eq.~\eqref{matrixelement}, are expected to obey the same scaling behavior, i.e., the $\mathcal{F}$ in Eq.~\eqref{dn} scales similarly to the left-hand side to Eq.~\eqref{ratio6} when $M> \Lambda$. 

While the prediction that at zero recoil the form factor scales like $1/M$ is fairly robust, and unsurprising to those familiar with HQET, one may be tempted to further conclude, based on these argument alone, that away from zero recoil that the differences in the slope of the form factors in $q^2$ at zero recoil would scale like $1/\Lambda$.  However, this is is not what is seen in some physical systems. For example, calculations on the lattice consistently claim that the $B_{(s)} \rightarrow D_{(s)}$ are only $\mathcal{O}(1\%)$ different over the entire kinematic range~\cite{Bailey:2012rr, Bailey:2014tva, Harrison:2017fmw, Monahan:2017uby, McLean:2019qcx}.  What is missing are the constraints from unitarity and analyticity, which are discussed in the following section.

\section{\normalsize Semileptonic $B \rightarrow D$ and $B_s \rightarrow D_s$ Decays}
\label{Bdecays}

The matrix element for semileptonic $B_{(s)} \rightarrow D_{(s)}$ transitions in the standard model, decomposed into a finite set of form factors, is:
\begin{equation}
\bra{ D_{(s)}(p') } \bar{c} \gamma^\mu b \ket{ B_{(s)}(p)} = (p+p')^\mu ~f^{(s)}_+(q^2) + (p-p')^\mu~f^{(s)}_-(q^2) \ee
\end{equation}
where $q = p - p'$.  For simplicity, we consider that the weak current is conserved, so $f^{(s)}_-$ does not contribute to the decay.  The constraints on the behavior of these form factors due to analyticity and unitarity were developed by the authors of Refs.~\cite{Meiman:1963, Okubo:1971jf, Okubo:1972ih, Singh:1977et, deRafael:1993ib}.  Such methods were utilized in Refs.~\cite{Boyd:1994tt, Boyd:1995cf, Boyd:1995sq, Boyd:1997kz} to not only constrain the behavior of the form factors that describe the semileptonic transitions ${B}\rightarrow D^{(*)}$, but also provide a parameterization of the form factors, relying on the remarkable fact that the entire kinematic range over which these particular transitions occur can be conformally mapped to a small analytic region within the unit disc.\footnote{Provided contributions from the $B_c \pi$ in the continuum are negligible~\cite{Boyd:1994tt, Boyd:1995cf, Boyd:1995sq, Boyd:1997kz}. }  This parameterization has been quite successful in extrapolating experimental data in order to determine the exclusive value of $|V_{cb}|$ at zero recoil~\cite{Grinstein:2017nlq, Bigi:2017njr}. The parameterization of the form factors developed by the authors of Refs.~\cite{Boyd:1994tt, Boyd:1995cf, Boyd:1995sq, Boyd:1997kz} is, very generically,
\begin{equation}
\label{BGLformfac}
f_k(z) =  \frac{1}{P_k(z)~\phi_k(z)} \sum_{n=0}^\infty a_n z^n \ee
\end{equation} 
and 
\begin{eqnarray}
\label{BGLdefs}
\sum_{n=0}^\infty |a_n|^2 \leq 1, \hspace{0.5in} z \equiv \frac{\sqrt{w+1}- \sqrt{2a}}{\sqrt{w+1}+ \sqrt{2a}}, \hspace{0.5in} w \equiv \frac{M_i^2 + M_f^2 - q^2}{2M_i M_f} \e
\end{eqnarray}
In Eq.~\eqref{BGLformfac}, the function $\phi_k(z)$ is known and depends on the specificities of the matrix element, the details of the chosen dispersion relation, and the unitarity bound, and $P_k(z)$ is a Blaschke factor where $|P_k(z)|=1$ when $|z|=1$, and chosen to be zero at the known location of any poles in the range $0\leq q^2 < (M_i + M_f)^{2}$.  
The factor $a$ in the definition of $z$ in Eq.~\eqref{BGLdefs} is a free parameter associated with what value of $z$ corresponds to what value of $q^2$.  Typically, $a=1$ is chosen, as done here, which corresponds to $z=0$ being the point of zero recoil, i.e., when $q^2 = (M_i - M_f)^2$.   Different choices of the value of $a$ can provide nominal improvement of the convergence of the Taylor expansion in Eq.~\eqref{BGLformfac} when fitting to data, but these this choice does not affect the discussion in this work.  
Importantly, the left-hand side of Eq.~\eqref{BGLformfac} is analytic for $|z|<1$, which justifies the Taylor expansion in $z$ on the right-hand side.  If considering the process where $X_i$ spontaneously transitions, (due to a local operator) into $X_f$, then $M_i > M_f$, $q^2\geq 0$, $0\leq z \leq z_\text{max}$, and  this expansion converges rapidly if
\begin{equation}
\label{smallz}
z_\text{max} \equiv \frac{\left( \sqrt{M_i} - \sqrt{M_f}\right)^4}{(M_i - M_f)^2} \ll 1 \e
\end{equation}
To illustrate, if $z_\text{max} \simeq 0.1$, then $M_f/M_i \simeq 0.27$, and  this corresponds to $(M_i-M_f)^2 / (M_i + M_f)^2 \lesssim 0.33$, which is not a small number compared to unity (one may note that this latter ratio of masses is known as the Shifman-Voloshin parameter~\cite{Shifman:1987rj}), which indicates that in this system, the smallness of the Shifman-Voloshin is not a necessary ingredient, and instead the constraints of unitarity and analyticity introduce a new small parameter:~$z_\text{max}$.

The form factors $f_+^{(s)}$ for semileptonic $B_{(s)} \rightarrow D_{(s)}$ decays can be parameterized as follows:
\begin{equation}
\label{BDformfac}
f_+(z) =  \frac{1}{P(z)~\phi(z)} \sum_{n=0}^\infty a_n z^n \ee \hspace{0.5in} \sum_{n=0}^\infty |a_n|^2 \leq 1 \ee
\end{equation} 
for $B\rightarrow D$, and 
\begin{equation}
\label{BsDsformfac}
f^{s}_+(y) =  \frac{1}{P(y)~\phi(y)} \sum_{n=0}^\infty b_n y^n \ee \hspace{0.5in} \sum_{n=0}^\infty |b_n|^2 \leq 1 \ee
\end{equation} 
for $B_s\rightarrow D_s$.  
The only differences between Eq.~\eqref{BDformfac} and Eq.~\eqref{BsDsformfac} is that $a_n \neq b_n$, in general, and one uses the masses of the $B$ and $D$ in Eq.~\eqref{BDformfac}, and the masses of the $B_s$ and $D_s$ in Eq.~\eqref{BsDsformfac}, i.e., $y$ has the same definition as $z$, but using the $B_s$ and $D_s$ masses in instead of the ones for $B$ and $D$. 
The masses of the poles and the unitarity bound are the same in both cases~\cite{Boyd:1994tt, Boyd:1995cf, Boyd:1995sq, Boyd:1997kz}. 
Here, $M_B \simeq 5.28$ GeV, $M_D \simeq 1.87$ GeV, $M_{B_s} \simeq 5.37$ GeV, $M_{D_s} \simeq 1.97$ GeV, and
these Taylor series converge very quickly, because Eq.~\eqref{smallz} is satisfied, where $z_\text{max} \simeq 6.4\%$ and $y_\text{max} \simeq 6.0\%$.  Expanding to first order can accurately parameterize the behavior of the form factors to a percent precision~\cite{Grinstein:2017nlq}:
\begin{equation}
\label{2formfacs}
f_+(z) \simeq  \frac{1}{P(z)~\phi(z)}  (a_0 + a_1z) \ee \hspace{0.5in} f^{s}_+(y) \simeq  \frac{1}{P(y)~\phi(y)} (b_0+b_1 y) \ee
\end{equation} 
where $a_0$, $a_1$, $b_0$, and $b_1$ are unknown constants, whose values can be determined by fitting to experimental or lattice data (the $a$'s here have no relation to those in Eq.~\eqref{ratio6}).

The quantum numbers for the $B$ and $B_s$ are the same, and likewise for the $D$ and $D_s$, except that their masses differ by $\mathcal{O}(1\%)$. 
Furthermore, the values of $z$ or $y$ over which the transitions $B\rightarrow D$ and $B_s\rightarrow D_s$ occur is known to be much smaller than the values of $z$ or $y$ at which the theory begins experiencing strong deformations, e.g., the locations of the $B_c$ poles or the $\bar{B}D$ threshold.  If so, one may expect that the result in Eq.~\eqref{ratio6} is applicable, since $M_B > M_D > \Lambda_\text{QCD}$.  To compare the two form factors, the differences in the definitions of $z$ and $y$ can be bounded from above by this ratio:
\begin{eqnarray}
\label{zy}
\frac{z_\text{max}}{y_\text{max}} \simeq 1 + \frac{2m_s}{\sqrt{M_B M_D}} + \mathcal{O}\left( m_s^2 \right)
\end{eqnarray}
where $m_s \simeq M_{B_s} - M_{B} \simeq M_{D_s} - M_{D}$. Therefore, one can let $z\simeq y$ for the sake of the scaling arguments that follow, and the error associated with this is subdominant, being at the sub-percent level.  Using the scaling argument at zero recoil (where $z=0$) in Eq.~\eqref{ratio6}:
\begin{eqnarray}
\label{a0scaling}
\frac{a_0}{b_0} &\simeq& 1 \pm \mathcal{O}\left( \frac{m_s}{M}  \right)  \\
&=& 1 \pm \mathcal{O}(1\%) \ee 
\end{eqnarray}
and away from zero recoil:
\begin{eqnarray}
\label{a1scaling}
\frac{a_1}{b_1} &\simeq& 1 \pm \mathcal{O}\left( \frac{m_s}{\Lambda_\text{QCD}}  \right) \ee \\
&=& 1 \pm \mathcal{O}(10\% - 50\%) \e
\end{eqnarray}
This means that the analogous form factors for $B\rightarrow D$ versus $B_s\rightarrow D_s$ should have approximately the same shape over the entire physical range of $q^2$, and differ in normalization by order of a few percent, because the relevant small parameter is $z_\text{max} m_s /\Lambda_\text{QCD} \simeq \mathcal{O}(1\%)$. 

These results can then be used to produce a parameterization of the ratio $R_s \equiv f_s(q^2)/f(q^2)$, which can be directly used $B\rightarrow D$ and $B_s \rightarrow D_s$ data, either from experiment or the lattice.  A derivation can be found in Appendix~\ref{Rsderivation}, for which the final result is:
\begin{eqnarray}
\label{Rs}
R_s \equiv \frac{f_s(q^2)}{f(q^2)}\simeq c_0 + c_1 y + \mathcal{O}(y^2)\ee
\end{eqnarray} 
where $c_0-1 \sim \mathcal{O}(m_s/M)$ and $c_1-1  \sim \mathcal{O}(m_s/\Lambda_\text{QCD})$. Because $R_s$ encapsulates the appropriate level of detail regarding the scaling arguments presented in this work, the linear expansion in the conformal variable on the left-hand side of Eq.~\eqref{Rs} is completely generic - it applies equally to all $B_{(s)}\rightarrow D^{(*)}_{(s)}$ form factors.
The result in Eq.~\eqref{Rs} is to be distinguished from the result using HQET alone (which neglects analyticity and unitarity), where, as discussed at the end of Section~\ref{formfactors}, one would expect this behavior around zero recoil:
\begin{eqnarray}
\label{onlyhqet}
R_s \simeq c_0 + \sum_{n=1}^\infty c_n (w(q^2)-1)^n \ee
\end{eqnarray} 
where $w$ is defined in Eq.~\eqref{BGLdefs}, which is essentially an expansion in $q^2$.  With HQET alone,  it is also expected that $c_0-1 \sim \mathcal{O}(m_s/M)$, however it is naively expected that $c_n-1  \sim \mathcal{O}(m_s/\Lambda_\text{QCD})$ for all $n\geq 1$.  

\begin{figure}
\centering     
\includegraphics[width=12cm]{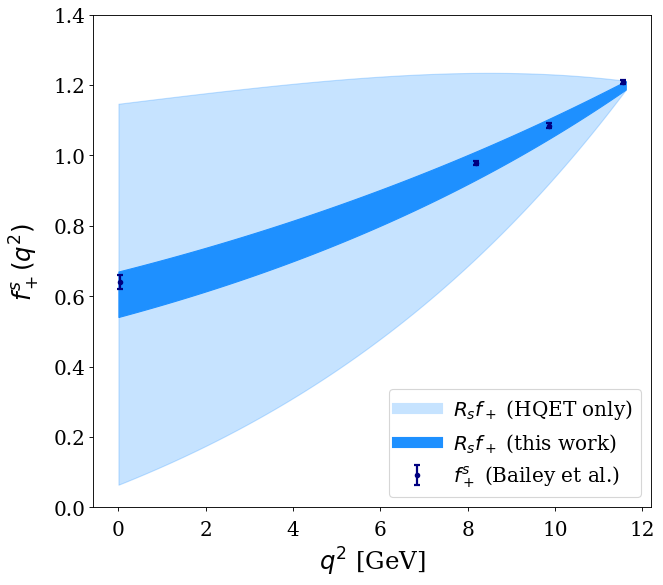}
\caption{The value of $f_+^s(q^2) \simeq R_s(q^2)f_+(q^2)$, where $f_+$ is a form factor associated with $B\rightarrow D$, for the parameterizations of $R_s$ using HQET only in Eq.~\eqref{onlyhqet} and those of this work in Eq.~\eqref{Rs}.  The interpolated function of $f_+(q^2)$ is used from Bailey {\it et al.}~(Ref.~\cite{Bailey:2012rr}), its uncertainties are ignored, and the results for $R_s(q^2)f_+(q^2)$ are shown in light and darker blue, with the parameterizations for $R_s$ in Eq.~\eqref{onlyhqet} and Eq.~\eqref{Rs}, varied by their nominal scales, respectively.  In this plot specifically, the HQET-only parameterization in Eq.~\eqref{onlyhqet} is truncated at linear order in the expansion, and $c_0$ and $c_1$ are varied between $1\pm 0.01$ and $1\pm 0.5$, respectively, in the two parameterizations of $R_s$ in Eq.~\eqref{onlyhqet} and Eq.~\eqref{Rs}.  The results from Ref.~\cite{Bailey:2012rr} for $f_+^s(q^2)$ are also shown with dark blue points with corresponding uncertainties.  While the differences between $f_+(q^2)$ and $f_+^s(q^2)$ are consistent with HQET alone, it is clear that there is additional relevant information in the system, i.e., that from analyticity and unitarity, which constrain the further differences between $f_+(q^2)$ and $f_+^s(q^2)$, and are the primary result of this work.    }
\label{fig1}
\end{figure}

In order to clearly compare the expectations from HQET alone, i.e., Eq.~\eqref{onlyhqet}, and the results of this work, i.e., Eq.~\eqref{Rs}, one can choose a particular form factor for $B\rightarrow D$ calculated on the lattice, multiply by the parameterization of $R_s$ in Eq.~\eqref{Rs} or Eq.~\eqref{onlyhqet}, and the result can be directly compared to a lattice calculation for the corresponding form factors for $B_s\rightarrow D_s$.  For example, we can use the interpolated value of $f_+(q^2)$ calculated on the lattice in Ref.~\cite{Bailey:2012rr}, multiply by $R_s$ either in Eq.~\eqref{Rs} or Eq.~\eqref{onlyhqet}, vary the $c_n$'s by their appropriate amounts, and compare with the values of $f_+^s(q^2)$ calculated in the same work.    
Fig.~\ref{fig1} shows the drastic differences in expectation between the scaling expectation in Eqs.~\eqref{Rs} and compared to the HQET-only expectation in Eq.~\eqref{onlyhqet}. 
Not shown is the positive consistency with other lattice results~\cite{Bailey:2012rr, Bailey:2014tva, Harrison:2017fmw, Monahan:2017uby, McLean:2019qcx}.  Of course, similar arguments holds for $B_{(s)}\rightarrow D^{*}_{(s)}$, and only one form factor comparison is shown here for simplicity.  It is clear from lattice data that $R_s(q^2)$ scales like the parameterization Eq.~\eqref{Rs}. 

The first version of this present work appeared in Oct.~2019, and since then the form factors for $B_s \rightarrow D^{(*)}_s \ell \nu$  have been measured by the LHCb experiment~\cite{Aaij:2020hsi}, though the data is still dominated by statistical uncertainties, so comparing the form factors for $B\rightarrow D^{*}$ compared to $B_{s}\rightarrow D^{*}_{s}$ is not yet possible at the percent level.  Also recently, a state-of-the-art result from the lattice was reported in Ref.~\cite{Lytle:2020tbe}, which shows clearly that the form factors for $B\rightarrow D$ differ from those for $B_s\rightarrow D_s$ at the percent level, over the entire kinematic range of the semileptonic decay.  These results from experiment and the lattice provide a robust confirmation of the scaling arguments in nonperturbative QCD presented here.

\section{\normalsize Summary}
\label{conclusions}

An assumption of continuity is that when the masses of massive external states are varied by a small amount $\varepsilon$, the form factors that parametrize the scattering process should also change by small amounts, as long as the process is far away from any strong deformations in the theory.  Because the form factors are dimensionless functions, the changes can either scale like $\varepsilon/M$ or $\varepsilon/\Lambda$, where $M$ is a typical mass scale in the process, and $\Lambda$ is the energy scale of the interactions.  

When the masses of the external scattering states are larger than $\Lambda$, and both the initial and final states have one heavy particle, then corrections away from the $M\rightarrow\infty$ limit are constrained to take on a very particular form, i.e., corrections at zero recoil scale like $\varepsilon/M$ and corrections away from zero recoil scale like $\varepsilon/\Lambda$ in $q^2$ space.  A discussion of this result is presented in Section~\ref{formfactors} and the derivation is presented in Appendix~\ref{app}, which is consistent with conclusions drawn from explicit calculations studying chiral symmetry breaking in HQET~\cite{Jenkins:1990jv, Boyd:1995pq}, despite the fact that the results in Section~\ref{formfactors} are make no mention of chiral symmetry or a Lagrangian.
When combined with the stringent constraints from unitarity and analyticity, as developed by the authors of Refs.~\cite{Boyd:1994tt, Boyd:1995cf, Boyd:1995sq, Boyd:1997kz}, the result is that away from zero recoil, corrections do indeed scale like $\varepsilon/\Lambda$, {\it but in $z$ space, not in $q^2$ space}. When applied to $SU(3)$ breaking in semileptonic $B_{(s)}\rightarrow D_{(s)}$ transitions, there is an accidentally small number, $z_\text{max}m_s/\Lambda \sim \mathcal{O}(1\%)$, and this means that the semileptonic $B \rightarrow D$ and $B_s \rightarrow D_s$ should differ by more than $\mathcal{O}(1\%)$ over the entire kinematic range, not just at zero recoil.  This conclusion is consistent with recent lattice calculations and distinguishable from the HQET-only expectation that the slope in $q^2$ of the form factors for semileptonic $B$ decays at zero recoil scale like $\varepsilon/\Lambda_\text{QCD}$, as can be seen in Fig.~\ref{fig1}.  

These results can provide valuable motivation for experimental analyses that combine results from $B\rightarrow D^*$ and $B_s\rightarrow D_s^*$ for a measurement of $|V_{cb}|$, or for lattice calculations to further confirm this non-trivial expectation in nonperturbative QCD.

\acknowledgements
This work is funded in part by the US DOE Office of Nuclear Physics and by the LDRD program at Los Alamos National Laboratory.  AK is grateful for useful feedback from Tanmoy Bhattacharya, Wouter Dekens, Benjam\'{i}n Grinstein, Andrew Lytle, John McGreevy, Emanuele Mereghetti, and Varun Vaidya, and for the hospitality of the physics department at UC San Diego.

\appendix

\section{Derivation of scaling behavior in $M\rightarrow \infty$ limit}
\label{app}

In the limit that $M\rightarrow \infty$,  the momentum of an on-shell degree of freedom with mass $M$ is $p^\mu = Mv^\mu$, where $v^\mu$ is the 4-velocity, and $v^2 = 1$, in order to match with the classical expectation.  Such states of heavy degrees of freedom can be prepared with a well-defined velocity and position, which can be understood heuristically as~\cite{Jenkins:1990jv, Boyd:1995pq}:
\begin{equation}
\Delta x \cdot \Delta p \geq \hbar/2 \hspace{0.25in} \rightarrow \hspace{0.25in} \Delta x \cdot \Delta v \geq \mathcal{O}\left( \frac{1}{M}\right) \e
\end{equation}
In the limit that $\Lambda/M \ll 1$, one can define a state with precise velocity, so the support of the spatial degrees of freedom of the heavy state obey the equation of motion:
\begin{equation}
\label{eom}
\frac{d v^\mu}{dt} = 0 \e
\end{equation}
Transformations that preserve this equation of motion elucidate some of the symmetries of the system in question.  Eq.~\eqref{eom} is related to heavy-quark symmetry.  Such heavy particles $X$ are non-relativistic, so these states can be factorized between their spatial degrees of freedom and everything else, which can be stated in momentum space as:
\begin{equation}
\label{factorization}
\ket{X} = \ket{\bf p} \otimes \ket{\text{other}} \ee
\end{equation}
where ``other'' means anything other than the momentum degrees of freedom, which includes spin.  Because of this factorization, the momentum degrees of freedom can be treated as those in a free theory, so, at this point in the discussion, we can speak of creation and annihilation operators acting on the vacuum corresponding to single-particle states.   In a relativistic theory, the momentum eigenstates $\ket{\bf p}$ are typically normalized as $\ket{\bf p} = \sqrt{2E_{\bf p}} a_{\bf p}^\dagger \ket{0}$, but in the $M\rightarrow \infty$ limit, this becomes:
\begin{equation}
\label{states2}
\ket{\bf p} = \sqrt{2M} a_{\bf p}^\dagger \ket{0}, \hspace{0.25in} \bra{0} a_{\bf p'} a_{\bf p}^\dagger \ket{0}  = (2\pi)^3~ \delta^3({\bf p}' - {\bf p}) \e
\end{equation}
A heavy particle travels in a straight line through spacetime in the $M\rightarrow \infty$ limit; there are no interactions in the Hilbert space that can change the particle's mass or velocity.  So, there are only transition amplitudes in the forward direction:
\begin{equation}
\bra{X_f} \mathcal{O} \ket{X_i} = 2M ~ \bra{\text{other}_f} \mathcal{O} \ket{\text{other}_i}\ee
\end{equation}
which follows from Eqs.~\eqref{factorization} and~\eqref{states2}.  Because this is a matrix element, the momentum-conserving delta function has been stripped off, and can be reintroduced when performing the phase space integral.  The $M$ appearing in this equation is the physical mass of the particle.  Now consider introducing a second heavy particle in the spectrum with mass $M'$, where the underlying interactions do not turn one heavy particle into another.  Taking the ratio of single-particle transition amplitudes for the same external momentum:
\begin{equation}
\frac{\bra{X_f} \mathcal{O} \ket{X_i}}{\bra{X'_f} \mathcal{O} \ket{X'_i}} = \frac{M}{M'} \frac{\bra{\text{other}_f} \mathcal{O} \ket{\text{other}_i}}{\bra{\text{other}'_f} \mathcal{O} \ket{\text{other}'_i}} \ee
\end{equation}
Note here that the quantum numbers of $X_i$ may be different than $X_f$, and likewise with $X'_i$ and $X'_f$, but the support of the momentum degrees of freedom are the same.  In the special case where the quantum numbers of $X_i$ are equal to $X_i'$ and those of $X_f$ are equal to $X_f'$ (which is the case being considered henceforth), then: 
\begin{equation}
\label{ratio}
\frac{M'}{M} \frac{\bra{X_f} \mathcal{O} \ket{X_i}}{\bra{X'_f} \mathcal{O} \ket{X'_i}} = 1 \e
\end{equation}
This ratio is trivially 1 if $M=M'$.  Eq.~\eqref{ratio} is the same expectation as in HQET~\cite{Shifman:1987rj, Isgur:1989vq,Isgur:1989ed,Manohar:2000dt, Jenkins:1992qv, Boyd:1995pq}.  
Note that in the limit that both $M\rightarrow \infty$ and $M' \rightarrow \infty$, that there is no additional hierarchy induced, e.g., neither $M/M'$ nor $M'/M$ become large in those limits.

Moving away from the $M, M' \rightarrow \infty$ limit, the interactions can change the velocity of the heavy particle, so neither the equation of motion in Eq.~\eqref{eom} nor the factorization in Eq.~\eqref{factorization} hold.  Again keeping the quantum numbers identical between initial and final states, the ratio in Eq.~\eqref{ratio} can only scale like the following:
\begin{equation}
\label{firstgoodeq}
\frac{M'}{M} \frac{\bra{X_f} \mathcal{O} \ket{X_i}}{\bra{X'_f} \mathcal{O} \ket{X'_i}} \sim 1+ \left( A + B ~ \delta v \right)  \left(\frac{\Lambda}{M} - \frac{\Lambda}{M'} \right) + \mathcal{O}\left( \frac{1}{M^2}\right) \ee
\end{equation}
since the ratio on the right-hand side must be 1 in the limit when $M, M'\rightarrow \infty$ or when $M = M'$.  Here, $A$ and $B$ are some constants  and $\delta v$ is a number directly proportional to the change in the velocity between the initial and final states, i.e., $\delta v$ need not be small, and when $\delta v$ is zero, this corresponds to there being no change in the velocity of the initial and final state particles. Continuity assumes that neither $A$ nor $B$ are $\gg \mathcal{O}(1)$.  Interestingly, if $M' = M+\varepsilon$, and $B=0$, then the right-hand side of Eq.~\eqref{firstgoodeq} scales like $\varepsilon/M$.

Now consider adding two more heavy particles, where there is another interaction, in addition to the one at scale $\Lambda$, which can turn heavy particles into other heavy particles, i.e., it allows the transition $X_i \rightarrow X_f$ and $X'_i \rightarrow X'_f$, as before, but now the masses of the states $X_i$, $X_f$, $X'_i$ and $X'_f$ are $M_i$, $M_f$, $M'_i$ and $M'_f$, respectively.  Even in the limit that all of these masses are infinite, this new interaction can insert momentum $q^\mu$, which can give rise to a change in the velocity of the heavy-particle trajectory, so the equation of motion  in Eq.~\eqref{eom} is not always conserved, i.e., one is moving away from a region in the phase space protected by heavy-quark symmetry.  However, at the special point where $q^{(\prime)}_\text{max} = v_i^{(\prime)}(M_i^{(\prime)} - M_f^{(\prime)})$, then $v_i^{(\prime)} =v_f^{(\prime)}$, and  the equation of motion in Eq.~\eqref{eom} maintains its form through the $X_i^{(\prime)} \rightarrow X_f^{(\prime)}$ transitions, and heavy-quark symmetry holds.  This point is called zero recoil, at which the ratio of amplitudes should have the same form as Eq.~\eqref{ratio}, since the spatial equation of motion is preserved:
\begin{equation}
\label{ratio3}
\frac{\sqrt{M'_i M'_f}}{\sqrt{M_i M_f}} \frac{\bra{X_f} \mathcal{O}(q_\text{max}) \ket{X_i}}{\bra{X'_f} \mathcal{O}(q'_\text{max}) \ket{X'_i}} = 1 \e
\end{equation}
The value of this ratio is the one calculated in the HQET literature~\cite{Shifman:1987rj, Isgur:1989vq,Isgur:1989ed,Manohar:2000dt}. 
Moving away from the zero recoil point by a small amount, the ratio scales like the following:
\begin{equation}
\label{ratio4}
\frac{\sqrt{M'_i M'_f}}{\sqrt{M_i M_f}} \frac{\bra{X_f} \mathcal{O}(q) \ket{X_i}}{\bra{X'_f} \mathcal{O}(q) \ket{X'_i}} \sim 1 + C~\delta v \left(\frac{M_i - M'_i}{\Lambda} + \frac{M_f - M'_f}{\Lambda} \right) + \mathcal{O}\left(\frac{1}{M}\right) \ee
\end{equation}
where $C$ is a constant.\footnote{The scaling in Eq.~\eqref{ratio4} is consistent with the scaling in the calculation in Refs.~\cite{Jenkins:1992qv, Boyd:1995pq, Manohar:2000dt}, where the $SU(3)_V$ flavor breaking effects were calculated in HQET for the ratio of the leading-order semileptonic Isgur-Wise functions for $B\rightarrow D$ versus $B_s\rightarrow D_s$.  }  Now $\delta v$ can depend non-trivially on $q$.  The right-hand side of Eq.~\eqref{ratio4} must be equal to 1 when $\delta v =0$ since it must reduce to Eq.~\eqref{ratio3} in that limit, or when $M_i  = M'_i$ and $M_f  = M'_f$ since the left-hand of Eq.~\eqref{ratio4} side becomes 1.  Including $1/M$ corrections, the ratio takes the form:
\small
\begin{eqnarray}
\label{ratio5}
\frac{\sqrt{M'_i M'_f}}{\sqrt{M_i M_f}} \frac{\bra{X_f} \mathcal{O}(q) \ket{X_i}}{\bra{X'_f} \mathcal{O}(q) \ket{X'_i}} &\sim& 1  + \left( A + B ~ \delta v \right)  \bigg( \frac{\Lambda}{M_i} - \frac{\Lambda}{M'_i} + \frac{\Lambda}{M_f} - \frac{\Lambda}{M'_f} \bigg) \nonumber \\
&& ~~ + ~ C~\delta v \left(\frac{M_i - M'_i}{\Lambda} + \frac{M_f - M'_f}{\Lambda} \right) \left( 1 + D \frac{\Lambda}{M_i} + E \frac{\Lambda}{M'_i} + F \frac{\Lambda}{M_f} + G \frac{\Lambda}{M'_f}\right) \nonumber \\
&& ~~ + ~ \mathcal{O}\left(\frac{1}{M^2}\right) \ee
\end{eqnarray}
\normalsize
since it must reduce to Eq.~\eqref{ratio4} as $M\rightarrow \infty$, and the right-hand side of Eq.~\eqref{ratio5} must be equal to 1 when $M_i = M'_i$ and $M_f = M'_f$, since in that case the right-hand side of Eq.~\eqref{ratio5} becomes 1.

This scaling argument does not rely on the fact that the $\delta v$ is small; its purpose is to show that there is a difference in scaling away from zero recoil.  One could replace any individual term proportional to $\delta v$ in the above equation with its own function, which depends on $v\cdot v'$, which goes to zero when $v\cdot v' = 1$.  Making the small differences in masses explicit, let $M'_i = M_i + \varepsilon$, and $M'_f = M_f + \delta$, and the ratio of matrix elements in Eq.~\eqref{ratio5} can be written as:
\small
\begin{eqnarray}
\label{ratio7}
\frac{\bra{X_f} \mathcal{O}(q) \ket{X_i}}{\bra{X'_f} \mathcal{O}(q) \ket{X'_i}} &\sim& 1 + a_1 \frac{\varepsilon}{M_i} + a_2 \frac{\delta}{M_f} + a_3\big(1+\mathcal{\chi}_1(v\cdot v')\big) \left( \frac{\Lambda}{M_i} \frac{\varepsilon}{M_i} + a_4 \frac{\Lambda}{M_f} \frac{\delta}{M_f} \right) \nonumber \\
&& ~~ + ~ \mathcal{\chi}_2(v\cdot v')  \left( \frac{\varepsilon}{\Lambda} + a_5 \frac{\delta}{\Lambda} \right) \left(1 + a_6 \frac{\Lambda}{M_i} + a_7 \frac{\Lambda}{M_f} \right) +  \mathcal{O}\left(\frac{\Lambda^2}{M^2},~ \frac{\varepsilon^2}{M^2},~ \frac{\delta^2}{M^2} \right) \ee
\end{eqnarray}
\normalsize
This is the result quoted in Eq.~\eqref{ratio6}.

\section{Derivation for $R_s$}
\label{Rsderivation}

Starting with the expressions for the form factors in Eq.~\eqref{2formfacs} and taking their ratio:
\begin{eqnarray}
R_s \equiv \frac{f_s(q^2)}{f(q^2)} = \frac{P(z)~\phi(z) \left(b_0 + b_1y \right)}{P(y)~\phi(y) \left( a_0 + a_1z \right)}\e
\end{eqnarray}
The presence of the $P(z)\phi(z)$ and $P(y)\phi(y)$ terms are due to the non-trivial constraints from unitarity, though the justification for truncating the Taylor expansion $P(z)\phi(z)f(z)$ and $P(y)\phi(y)f(y)$ at linear order also applies to the ratio $R_s$:
\begin{eqnarray} 
\label{Rs1}
R_s = d_0 + d_1 z + d_2 y + \mathcal{O}\left(z_\text{max}^2, ~y_\text{max}^2, ~z_\text{max}y_\text{max} \right)\ee
\end{eqnarray}
where $d_0$, $d_1$, and $d_2$ are constants, known functions of $a_0$, $a_1$, $b_0$, $b_1$, $M_B$, $M_D$, $M_{B_s}$, $M_{D_s}$, and the $M_{B_c}$ pole masses contained in the definition of the Blaschke factor $P$.  One can check via a straightforward calculation that if Eqs.~\eqref{a0scaling} and~\eqref{a1scaling} are true, and if $a_1/a_0$ (or equivalently $b_1/b_0$) is not $\gg 1$, then the Taylor series in Eq.~\eqref{Rs1} is justified for the physical values of the mesons (meaning the values of $d_0$, $d_1$, and $d_2$ are not $\gg 1$).   In particular, an important and general expectation is:
\begin{eqnarray}
\label{bigresult}
d_0 = 1 + \delta, \hspace{0.25in}\text{where} ~ ~\delta \sim \mathcal{O}\left(\frac{m_s}{M} \right)\ee
\end{eqnarray}
and
\begin{eqnarray}
d_1, d_2 = 1 + \Delta , \hspace{0.25in}\text{where} ~ ~\Delta \sim \mathcal{O}\left(\frac{m_s} {\Lambda_\text{QCD}} \right) \ee
\end{eqnarray}
These results are derived using Eqs.~\eqref{a0scaling} and~\eqref{a1scaling}, respectively.  

To illustrate the result in Eq.~\eqref{bigresult} is correct with an explicit choice of the form factors associated with $B_{(s)} \rightarrow D_{(s)}^{(*)}$, we can choose the form factor $f_{(s)}$ for $B_{(s)}\rightarrow D_{(s)}^*$, as defined in Refs.~\cite{Boyd:1997kz, Grinstein:2017nlq}, from which one can calculate $d_0$ explicitly:
\begin{eqnarray}
\label{d0ex}
d_0 = \frac{M_{B_s}^2}{M_B^2}\frac{b_0}{a_0} \frac{r\left(1+2\sqrt{r_s} + r_s \right)^4}{r_s\big(1+\sqrt{r} \big)^8} \prod_{i=1}^4 \frac{z_{P_i}}{y_{P_i}} \ee
\end{eqnarray}
where $r \equiv M_D/M_B$, $r_s \equiv M_{B_s}/M_{D_s}$, and $z_{P_i}$ and $y_{P_i}$ are the locations of the $B_c$ poles in $z$ and $y$ space, respectively.  Using the results in Eqs.~\eqref{zy} and \eqref{a0scaling}, it is then clear that Eq.~\eqref{d0ex} is consistent with Eq.~\eqref{bigresult}. The same expectation is also true for all the other form factors in $B_{(s)} \rightarrow D_{(s)}^{(*)}$.   The same can be done with $d_1$ and $d_2$, though the calculation is considerably more complicated and will not be illustrated here for the sake of brevity. 

Continuing with the parameterization for $R_s$, using the fact that Eq.~\eqref{zy} is an upper bound on the differences between $z$ and $y$ across the entire kinematic region,
\begin{eqnarray}
R_s \simeq c_0 + c_1 y \e
\end{eqnarray} 
Here, $y$, and not $z$, is chosen, because the there are values of $q^2$ in $B \rightarrow D$ that are beyond the kinematic limits of $B_s\rightarrow D_s$, allowing for a comparison between the form factors in $q^2$.  Again, $c_0 -1 \sim \mathcal{O}(m_s/M)$ and $c_1 -1 \sim \mathcal{O}(m_s/\Lambda_\text{QCD})$.  This is the result stated in Eq.~\eqref{Rs}.

\bibliographystyle{JHEP}

\bibliography{bib}{}

\end{document}